\documentstyle[12pt]{article}
\begin{document}
\renewcommand{\floatpagefraction}{.85}
\begin{center}
\vspace*{0.5in}
{\LARGE Direct mechanism in\\
\vspace{12pt}
solar nuclear reactions}\\
\vspace{20pt}
{\large H.~Oberhummer$^{*}$ and G.~Staudt}$^{**}$\\
\vspace{20pt}
{\small \it $^{\it *}$Institut f\"ur Kernphysik, TU Wien,
A--1040 Vienna, Austria and\\
$^{**}$Physikalisches
Institut der Univ.~T\"ubingen, D--72706 T\"ubingen, Germany}
\end{center}
\vspace{12pt}
\begin{quote}
{\footnotesize {\bf Abstract.} A short overview of the direct reaction
mechanism
and the models used for the analysis of such processes is given.
Nuclear reactions proceeding through the direct mechanism and
involved in solar hydrogen burning are discussed.
The significance of these nuclear reactions
with respect to the solar neutrino problem is investigated.}
\end{quote}
\vspace{6pt}
\begin{center}
{\large {\bf INTRODUCTION}}
\end{center}
\vspace{2pt}

In the last years the importance of the direct reaction (DI)
mechanism in stellar nuclear reactions has been realized.
This mechanism is of relevance for many nuclear reactions in
primordial and stellar nucleosynthesis. Obviously,
the best stellar data are obtained from the
observation of the sun. In this article we want to concentrate
on direct nuclear reactions taking place in the pp--chain of solar
hydrogen burning and being related to the solar neutrino problem; i.e.~that the
number of solar neutrinos observed by earth--bound detectors
is significantly less than predicted by the Standard Solar Model (SSM).

This article concentrates more on solar reactions rates obtained from
calculational
efforts than from experimental determination.
In section 2 we give a short overview of the DI mechanism and
models. These models are {\it microscopic methods\/},
as the Resonating Group Method (RGM) and Generator Coordinate Method (GCM)
and {\it potentials models\/}, as the
Optical Model (OM), Distorted Wave Born Approximation
(DWBA) and Direct Capture (DC).
In section 3 we discuss some nuclear reactions which proceed predominately
through the direct reaction mechanism and are of interest
for the solar neutrino problem. In the following section we discuss
some nuclear aspects of the solar neutrino problem.
Finally, in the last section we give a summary.
\vspace{6pt}
\begin{center}
{\large {\bf DIRECT REACTION MECHANISMS AND MODELS}}
\end{center}
\vspace{2pt}

In nuclear reactions two extreme types of reaction mechanisms can exist:
The compound--nucleus (CN) and direct (DI) process. In the CN mechanism
the projectile merges in the target
nucleus and excites many degrees of freedom of the CN. The excitation proceeds
via a multistep process and therefore has a reaction time typically of the
order $10^{-16}\,$s to $10^{-20}\,$s. After this time the CN decays into
various exit channels. The relative importance of the decay channels is
determined by the branching ratios to the final states.
In the DI process the projectile excites only a
few degrees of
freedom (e.g.~single--particle or collective). The excitation proceeds in one
single step and has a characteristic time scale of $10^{-21}\,$s to
$10^{-22}\,$s. This corresponds to the time the projectile needs
to pass through the target nucleus; this time is much shorter than the
reaction time of CN
processes. However, at subCoulomb energies the reaction is now hindered by the
Coulomb
and centrifugal barriers. Therefore, the characteristic time scale
is enhanced by a factor determined by the barrier penetration
pro\-ba\-bilities.

The question whether a given reaction favours CN or DI processes depends
on the reaction considered and on the relative energies in
the entrance and exit channel. In general, a given reaction involves
both types of reaction mechanisms and also intermediate types
(e.g. precompound reactions). However, for certain reactions
and projectile energy ranges one type of reaction mechanism may dominate.

In thermonuclear scenarios the projectile energy is well below the Coulomb
and/or centrifugal barrier. At these energies the competition between different
reaction
mechanisms is quite complicated. At these
energies the CN formation may be suppressed, because there may exist no
CN levels that can be populated, especially in light or magic nuclei.
If the  overlap between entrance-- and exit--channel wave functions is large,
then
the DI process will be enhanced. This may be the case if the
wave function of the transferred nucleon or nucleon cluster is weakly bound.
Then the reactions may take place well outside the nucleus. On the other hand
in this case the CN process is suppressed, because the overlap
with the confined bound--state type wave functions which are responsible for
the
CN process, is small.

Closely connected to the DI and CN mechanisms are the resonance types
in the nuclear reaction cross sections. There are again two extreme cases for
resonance structures. The first type (compound resonance) occurs in CN
reactions
and has a
narrow width (order of $1$ to $100\,$eV). This width is
determined by the long reaction time corresponding to the many steps
involved in the formation and decay of the CN.
The second type is called  potential, optical, single--particle,
poten\-tial-well or
scattering resonance. This resonances type is associated with DI and is
determined  by the optical potential
in the entrance or exit channel. Usually
the widths of potential resonances
are broad (order of 10 keV
to a few MeV) corresponding to the short reaction times of the DI. These
resonances lose their meaning for
higher energies because of their broad widths, as a result of absorption
into other channels. However, for energies below the Coulomb and centrifugal
barriers the width of potential resonances may become very narrow
due to increasing lifetimes.

First--principle microscopic theories, such as the resonating
group method (RGM) or generator coordinate method (GCM),
are based on many--nucleon wave functions of the nuclei involved and
on nucleon--nucleon interactions
(1), (2). 
In this approach, the explicit inclusion of the Pauli principle leads to
complicated highly nonlocal potentials for the interaction between
the composite nuclei in the entrance and exit channel.

The main drawback of the RGM is that it requires extensive analytical
calculations without systematic character when going from one reaction
to another. Consequently, the application of the RGM is
essentially restricted to reactions involving only a small number
of nucleons. This problem can be overcome by the GCM which
is similar to the RGM, but it allows systematic
calculations, well adopted to a numerical approach.
In the GCM the relative wave functions
are expanded in a Gaussian basis. Nowadays, most
microscopic calculations are performed in
the GCM framework.

It is obvious that a fully microscopic approach like RGM and GCM is
more satisfying, since it is a first--principle approach. Such a microscopic
model starts from the nucleon--nucleon interaction and does not
contain any free parameters. It is therefore possible to predict
physical properties of the system independently of experimental
data. However, in most cases the fully microscopic approach rarely
reproduces physical quantities which are fundamental for the calculation of
astrophysical cross sections, such as thresholds or resonance
energies and their widths. Nevertheless, such methods
have been used in the description of astrophysically relevant processes by
allowing adjustments of the nucleon--nucleon interaction parameters,
by renormalizing bound and resonance energies or astrophysical S--factors.
Review articles on microscopic theories and/or their application
to astrophysical processes are found in
(3)--(7).

The potential model is based on the
description of the dynamics of nuclear processes by a Schr\"odinger
equation with local potentials in the entrance and exit channels
and is appropriate for the description of direct reactions.
Potential models can be applied to elastic scattering
(Optical Model: OM), to transfer reactions (Distorted
Wave Born Approximation: DWBA) and to capture reactions
(Direct Capture: DC). The most important ingredients of
potential models are the wave functions for the scattering,
bound and quasi--bound (resonant) states in the entrance and
exit channels. In the work performed by our group these wave functions
have been determined from potentials calculated by the folding procedure
(5), (8), (9). The nuclear densities for the folding procedure are
derived from nuclear charge distributions with an energy-- and
density--dependent NN--interaction. Important for the success of the
potential models is the fact that the strength of the potentials
are fitted to experimental elastic scattering cross sections and to the
energies of bound and quasi--bound states. In this respect the potential models
together with the folding procedure
combine the first--principle approach of a microscopic theory with the
flexibility of a phenomenological method.
\vspace{6pt}
\begin{center}
{\large {\bf NUCLEAR REACTIONS IN SOLAR HYDROGEN BURNING}}
\end{center}
\vspace{2pt}

In this section we want to discuss calculations of reactions taking
place in solar
hydrogen burning. We concentrate on reactions determining the
branching ratios of the pp--chain, which are of interest
for the solar neutrino problem. These reactions proceed predominately
through the direct reaction mechanism. The main solar neutrino flux
(95\% for the $^{37}$Cl experiment, 100\% for the Kamiokande
experiment and 92.5\% for the $^{71}$Ga experiments)
come from the three pp--chains (Figs.~1 and 2).
The nuclear reactions which determine the branching ratios of the pp--chain
are the following (Fig.~1):
\begin{enumerate}
\item The reactions
$^{3}$He($^{3}$He,2p)$^{4}$He
and $^{3}$He($\alpha$,$\gamma$)$^{7}$Be
determine the branching ratio
between the ppI-- and (ppII+ppIII)--chains.
\item The reactions
$^{7}$Be(p,${\gamma}$)$^{8}$B and
$^{7}$Be(e$^{-}$,${\nu}$)$^{7}$Li determine the branching ratio
between the ppII-- and ppIII--chains.
\end{enumerate}
\begin{figure}
\vspace{8.5cm}
\begin{center}
{\small {\bf FIGURE 1.}~The pp--chain (adapted from (10)).}
\end{center}
\end{figure}
\begin{figure}
\vspace{9.3cm}
{\small {\bf FIGURE 2.}~The neutrino flux as a function of the
neutrino energy from the different neutrino
reactions in solar hydrogen burning. On the upper scale
also the thresholds of the earth--bound neutrino
experiments are shown (adapted from (10)).}
\end{figure}

A general overview of the solar neutrino problem up to 1988
containing also the relevant nuclear processes can
be found in the book of Bahcall (11).
The main emphasis of this article
is on recent theoretical calculations of astrophysical
S--factors for the above reactions and their comparison with the currently
accepted low--energy values (12) and recent experimental data.
\vspace{6pt}
\begin{center}
{\large {\bf The reactions $^{\bf 3}$He($^{\bf 3}$He,2p)$^{\bf 4}$He
and $^{\bf 3}$He(${\bf \alpha}$,${\bf \gamma}$)$^{\bf 7}$Be}}
\end{center}
\vspace{2pt}
Since the reactions $^{3}$He($\alpha$,$\gamma$)$^{7}$Be
and $^{3}$He($\alpha$,$\gamma$)$^{7}$Be are important for determining the
branching
ratio of the ppI-- and (ppII+ppIII)--chains in solar hydrogen burning, the
magnitudes of the
cross sections for both reactions are of special interest for the solar
neutrino problem
(11).

The reaction $^{3}$He($^{3}$He,2p)$^{4}$He has been measured by a number
of authors in the subCoulomb energy range (13)--(16), most recently by Krauss
et
et al.~(17). In the later work the astrophysical S--factor was measured
in the energy range $E_{\rm c.m.}=17.9$--$342.9$\,keV. These data extend into
the
the thermonuclear energy region of the sun. In agreement with other
experimental results ((18) and references therein) no evidence for a suggested
low--energy resonance (19), (20) has been found. Such a low--energy resonance
would significantly
enhance the $^{3}$He+$^{3}$He--route in the pp--chain (21).

{}From the above measurements
the data has been extrapolated to the thermonuclear energy range using either
phenomenlogical models with quadratic polynomials (17), microscopic methods
like the coupled--channel RGM (22) or the potential model (23) (Fig.~3).

In the RGM (22) as well as the DWBA (23) calculation the exit channel of the
reaction $^{3}$He($^{3}$He,2p)$^{4}$He is treated by the formation
of an $\alpha$--particle and a diproton cluster. In
both calculations parameters are adjusted to properties
of the combined cluster sytem in such a way that it fits
reasonably the low--energy experimental reaction data.
\begin{figure}
\vspace{10cm}
{\small {\bf FIGURE 3.}~Astrophysical S--factor for the reaction
$^{3}$He($^{3}$He,2p)$^{4}$He calculated from
a microscopic calculation (22)
(short--dashed curve) and
the potential model (23) (long--dashed curve) and compared to the
currently accepted value low--energy values (12) (solid line) and experimental
data [circles (15)], [triangles (17)].}
\end{figure}

Experimental data for the $^{3}$He(${\alpha}$,${\gamma}$)$^{7}$Be
cross section at subCoulomb energies has been obtained
by numerous experimental groups (24)--(29). Capture cross
sections observing the decay of $^{7}$Be residual
nucleus have been measured in further experiments (27), (29)--(31).
The experimental data of Kr\"ahwinkel et al.~(26) have been renormalized later
on by a factor of 1.4 as suggested by Hilgemeier et al.~(29).

Theoretically, Tombrello and Parker (32) have first succeeded
in describing the energy dependence of the astrophysical
S--factor in a direct--capture model using a
hard--sphere potential. Further calculations in the framework of the
potential model have been carried out in (33).
The scattering wave functions of $^{3}$He and $^{4}$He
and the bound--state wave functions of $^{7}$Be were
constructed by a phenomenological Woods--Saxon potential
model and the orthogonality condition model. These models
account for the measured elastic scattering and excitation
energies of the low--lying states of $^{7}$Be. In a further approach
(34) for the lowest states in A=7 nuclei the model
of a real $^{3}$He nucleus interacting with a
$\alpha$--particle through a deep local potential of
Gaussian shape is adopted. The four parameters of the
model are determined by fitting them to reproduce numerous
independent data in $^{7}$Be.

Analyses using microscopic theories based on the RGM have been
performed in (35)--(40). In the work of Kajino and arima(36) no normalization
is used in order to obtain the astrophysical S--factor. The
Pauli principle is fully taken into account. The radiative
capture and the elctromagnetic properties  using a multichannel RGM
which also allows for studying the influence of inelastic scattering
and distortion effects has been investigated in (37). For the
relative cluster motion a number of Gaussian functions with
different width parameters have been used. The capture
reaction $^{3}$He(${\alpha}$,${\gamma}$)$^{7}$Be was also studied
in the framework of a microscopic potential model which is justified on the
basis
of microscopic many--body scattering theories (38).

In our calculation of the reactions $^{3}$He($^{3}$He,2p)$^{4}$He
and $^{3}$He(${\alpha}$,${\gamma}$)$^{7}$Be in the potential
model the folding potentials are adjusted to reproduce
the differential cross sections of the $^{3}$He--$^{3}$He
and $^{3}$He--$^{4}$He elastic scattering measured in
the range 1--3\,MeV. The fit to the cross section data
results in a strong parity dependance of the potentials.
The $^{3}$He--$^{3}$He potentials also describe very well
the elastic scattering phase shift calculated in the
RGM model ((23) and references therein) up to an energy
of $E_{\rm lab} = 30$\,MeV. In the $^{3}$He--$^{4}$He
case good agreement is found between the measured cross sections
and the experimental and calculated elastic phase shifts
up to $E_{\rm lab} \approx 12$\,MeV ((41) and references therein).

The optical potentials obtained were used to calculate the astrophysical
S--factors of the transfer reaction $^{3}$He($^{3}$He,2p)$^{4}$He
and the capture reaction $^{3}$He(${\alpha}$,${\gamma}$)$^{7}$Be.
For the diproton system we choose a Fourier--Bessel
charge--distribution. The strenght of the $^{4}$He--2p potential
was adjusted to reproduce the experimental reaction cross section
at 150\,keV. The spectroscopic factors for the above calculations
were taken from shell--model calculations.

In Figs.~3 and 4 the results of our calculations are given by
the long--dashed curves. For both reactions the agreement between
the experimental and calculated data is excellent. The $S_{0}$--factors
obtained in our calculations agree very well with the currently accepted values
(12) if the experimental errors are taken in consideration (Table 1).
\begin{figure}
\vspace{10cm}
{\small {\bf FIGURE 4.}~Astrophysical S--factor for the reaction
$^{3}$He(${\alpha}$,${\gamma}$)$^{7}$Be calculated from
different microscopic calculations (35)--(39)
(upper and lower limits are shown by short--dashed curves) and
the potential model (41) (long--dashed curve) and compared to the
currently accepted low--energy values (12) (solid line) and experimental
data [circles (26), triangles (27)].}
\end{figure}
\begin{table}
{\bf Table 1}.~Recently calculated low--energy astrophysical S--factors
for reactions of the pp--chain (upper part) compared to the currently accepted
values
of Bahcall (12) (lower part).
{\small
\begin{center}
\begin{tabular}{|c|c||c|c|}
\hline
Reaction & Reference & $S_{0}$ [keV\,b] & $dS/dE$ [b]\\
\hline
$^{3}$He($^{3}$He,2p)$^{4}$He & (23) &
$4.854 \cdot 10^{3}$ & $-1.328$ \\
$^{3}$He($\alpha$,$\gamma$)$^{7}$Be & (41) &
0.516 & $-3.67 \cdot 10^{-4}$ \\
$^{7}$Be(p,$\gamma$)$^{8}$B & (47)&
0.0249 & $-3.2 \cdot 10^{-5}$ \\
\hline
$^{3}$He($^{3}$He,2p)$^{4}$He & (12) &
$5.0(1 \pm 0.18) \cdot 10^{3}$ & $-0.9$\\
$^{3}$He($\alpha$,$\gamma$)$^{7}$Be & (12) &
$0.533(1 \pm 0.096)$ & $-3.1 \cdot 10^{-4}$\\
$^{7}$Be(p,$\gamma$)$^{8}$B & (12) &
$0.0224(1 \pm 0.28)$ & $-3 \cdot 10^{-5}$\\
\hline
\end{tabular}
\end{center}}
\end{table}

Compared to the microscopic calculations the potential--model
calculations are not dependent on different adopted model spaces and
effective nucleon--nucleon interactions. Instead the relevant scattering
and bound--state wave functions are determined uniquely from
elastic scattering data and resonant and bound--state energies
(except for the $^{3}$He--2p system, where no experimental data
is available). For the $^{3}$He($^{3}$He,2p)$^{4}$He reaction
the microscopic calculation (22) underestimates the data at
at higher energies, when compared to the experimental data (15), (17),
to the potential model calculation (23) and to the currently accepted
low--energy values (12) (Fig.~3). The reason for this deficiency
is probably that in the microscopic calculation only s-- and d--waves
in the entrance channel are included, whereas in the potential model
partial waves up to $\ell=6$ are taken into account.
For the $^{3}$He($\alpha$,$\gamma$)$^{7}$Be reaction the energy
dependance of the different microscopic and the potential--model
calculations are quite similar. However, for the absolute value
the astrophysical S--factor for the different microscopic calculations
scatter within about a factor of 2 (Fig.~4). This is mainly due
to the different adopted effective nucleon--nucleon interactions
in the microscopic calculations. Again
the potential--model gives an unique value, because the wave
functions are determined from the elastic scattering data and
resonant and bound--state energies. Contrary to the
potential--model only the lower limit of
the microscopic calculations can reproduce the
experimental data (26), (27) and the currently accepted
low--energy values (12) (Fig.~4).
\vspace{6pt}
\begin{center}
{\large {\bf The reactions $^{\bf 7}$Be({\bf p},${\bf \gamma}$)$^{\bf 8}$B
and $^{\bf 7}$Be({\bf e}$^{\bf -}$,${\bf \nu}$)$^{\bf 7}$Li}}
\end{center}
\vspace{2pt}

The most uncertain of all nuclear reaction rates relevant for
the solar neutrino problem is the reaction $^{7}$Be(p,${\gamma}$)$^{8}$B.
The reason is that there exist two measurements at low energies
which differ by about 30\% in the absolute magnitude for the
astrophysical S--factor (42), (43). Both
measurements find about the same low--energy dependance of
the astrophysical S--factor in accordance with the microscopic
calculations (44)--(46) and the potential--model
approach (47). However, due to different adopted model spaces and
effective nucleon--nucleon interactions also the absolute
values of the astrophysical S--factors for the
microscopic calculations differ considerably
(Fig.~5).
Therefore, the absolute values of the astrophysical factors
obtained from the microscopic calculations have either
been fitted to the two existing experimental data sets (45) or
a range of different possible values for the astrophysical S--factor
is presented by the authors (46).

In the potential--model approach for the reaction
$^{7}$Be(p,${\gamma}$)$^{8}$B the strength of the
folding potentials are adjusted to reproduce the
resonant and bound states in the entrance and exit channel (47).
This ensures that the asymptotic parts of the wave functions
are correct. This is of special importance, because
the main contributions for this reaction come from
far out in the nuclear exterior (for the reaction
$^{7}$Be(p,${\gamma}$)$^{8}$B at 15\,keV the
main contributions come from a region of about 40\,fm outside
the target nucleus) (47). The energy dependence
of the astrophysical factors is consistent with the
experimental data sets as well as the microscopic
calculations. However, using
the wave functions obtained with the folding procedure it is also possible
to calculate the absolute values of the astrophysical
S--factor for this reaction without using a renormalization
procedure (47). This calculation reproduces almost the data of
Kavanagh et al.~(42) (Fig.~5) and therefore favours the higher experimental
value of the astrophysical S--factor. The potential--model
calculation is about 11\% higher than the
currently accepted value (12) (Table 1 and Fig.~5).
\begin{figure}
\vspace{10cm}
{\small {\bf FIGURE 5.}~Astrophysical S--factor for the reaction
$^{7}$Be(p,${\gamma}$)$^{8}$B calculated from
different microscopic calculations (44)--(46)
(upper and lower limits are shown by short--dashed curves) and
the potential model (47) (long--dashed curve) and compared to the
currently accepted low--energy values (12) (solid line) and experimental
data [squares (42), diamonds (43)].}
\end{figure}

Recent progress has also been made on the electron
capture rate by a  $^{7}$Be--nucleus through
the reaction $^{7}$Be(e$^{-}$,${\nu}$)$^{7}$Li in the solar plasma
by performing a self--consistent study of continuum
and bound electrons (45). The improved treatment
of the screening effects results in a small increase
for this reaction rate
of about 1.3\% compared to the currently accepted value given in (11).
\newpage
\begin{center}
{\large {\bf NUCLEAR ASPECTS OF THE SOLAR NEUTRINO PROBLEM}}
\end{center}
\vspace{2pt}

The neutrino--producing nuclear reactions in the sun are shown
in Fig.~2. As can be seen from this figure the strongest neutrino
flux stems from the reaction ${\rm p}+{\rm p} \rightarrow {\rm d}+{\rm
e}^{+}+\nu$.
However, these neutrinos have very low energies $E_{\nu} \le 0.42$\,MeV.
On the other hand almost all the neutrinos with $E_{\nu} > 2$\,MeV
are produced in the decay of $^{8}$Be. The neutrino energies play
a crucial role in their detection, because all neutrino detectors
have certain thresholds (see upper scale in Fig.~2). The threshold
energies are 0.814\,MeV for the Davis ($^{37}$Cl--detector),
about 7.5\,MeV for the Kamiokande (neutrino--electron
scattering) and 0.233\,MeV for the Gallex and Sage experiments
($^{71}$Ga--detector).

The predicted neutrino flux of the SSM
is compared to the different observed solar neutrino rates
from the Davis, Kamiokande, Gallex and Sage experiments
in Table 2. Roughly speaking one can say that the observed neutrino flux
is about 1/3, 1/2 and 2/3 of the predicted neutrino flux for
the Davis, Kamiokande, and Gallex and Sage experiments, respectively.
\begin{table}
{\bf Table 2}.~Observed solar neutrino rates compared to the predictions
of the SSM. (From (12), (48)).
{\small
\begin{center}
\begin{tabular}{|c|c|c|c|}
\hline
Experiment & Observation (SNU)$^{\rm a}$ & Prediction (SNU)$^{\rm a}$ &
$\frac{\rm Observation}{\rm Prediction}$$^{\rm b}$\\
\hline
Davis & $2.2 \pm 0.2$ & $8.0 \pm 3.0$ & $0.275 \pm 0.025$ \\
Kamiokande &  &  & $0.47 \pm 0.05 \pm 0.06$ \\
Gallex & $83 \pm 19 \pm 8$ & $131.5^{+20}_{-17}$ &
$0.63 \pm 0.20$ \\
Sage & $58^{+17}_{-24} \pm 14$ & $131.5^{+21}_{-17}$ &
$0.44^{+0.24}_{-0.29}$ \\
\hline
\end{tabular}
\end{center}}
{\footnotesize
$^{\rm a}$ 1 SNU = 1 {\underline S}olar
{\underline N}eutrino {\underline U}{\rm nit\/} = 10$^{-36}$\,$\nu$--captures
per s and target atom.\\
$^{\rm b}$ The errors reflect only the observational errors.}
\end{table}
In the following we want to investigate the dependance of the
predicted fluxes for the different solar neutrino experiments
on the nuclear input using a simple model.
The fluxes $\phi$ for the different solar neutrino sources of
the pp--chain can be expressed through the
following proportionality relations
(assuming constant luminosity, radius, metal abundance and age of the sun)
(11):
\begin{equation}
\phi({\rm pp}) \propto \phi({\rm pep})
\propto S_{11}^{0.14}S_{33}^{0.03}S_{34}^{-0.06}
\end{equation}
\begin{equation}
\phi(^{7}{\rm Be}) \propto S_{11}^{-0.97}S_{33}^{-0.43}S_{34}^{0.14}
\end{equation}
\begin{equation}
\phi(^{8}{\rm B}) \propto S_{11}^{-2.6}S_{33}^{-0.40}S_{34}^{0.81}S_{17}^{1.0}
\tau_{{\em e}7}^{1.0}
\end{equation}
In these equations the $S$ denote the low--energy astrophysical S--factors
for the different reactions wich are characterized by:\\
pp: ${\rm p}+{\rm p} \rightarrow {\rm d}+{\rm e}^{+}+\nu$\\
pep: ${\rm p}+{\rm e}^{-}+{\rm p} \rightarrow {\rm d}+\nu$\\
33: $^{3}$He($^{3}$He,2p)$^{4}$He\\
34: $^{3}$He($\alpha$,$\gamma$)$^{7}$Be\\
17: $^{7}$Be(p,${\gamma}$)$^{8}$B\\
e7: $^{7}$Be(e$^{-}$,${\nu}$)$^{7}$Li.

Compared to the formulae given by Bahcall (11) we multiplied the reaction rate
for the $^{8}$B--neutrino flux additionally with the factor
$\tau_{{\rm e}7}$. This factor describes the lifetime of $^{7}$Be
with respect to the reaction $^{7}$Be(e$^{-}$,${\nu}$)$^{7}$Li.
Since the concentration of $^{7}$Be is proportional to
its lifetime with respect to electron capture of $^{8}$Be (the
lifetime with respect to $^{7}$Be(p,${\gamma}$)$^{8}$B
is much longer) the $^{8}$B--neutrino flux is proportional
to the lifetime $\tau_{{\rm e}7}$ as well as to $S_{17}$.

For the different neutrino--experiments we can deduce the
following expressions for the predicted solar neutrino fluxes in
the SSM for the different relevant reaction rates
of the pp--chain:
\begin{enumerate}
\item $^{37}$Cl experiment using (12):
\begin{eqnarray}
\Phi(^{37}{\rm Cl}) & = & (0.2 R({\rm pep}) + 1.2 R(^{7}{\rm Be}) +\\ \nonumber
& & + 6.2 R(^{8}{\rm B}) + 0.4)\,{\rm SNU}
\end{eqnarray}
\item Kamiokande experiment:
\begin{equation}
\frac{\Phi({\rm Kamiokande (Observation)})}
{\Phi({\rm Kamiokande (Prediction)})} =
0.47R(^{8}{\rm B})
\end{equation}
\item $^{71}$Ga experiments:
\begin{eqnarray}
\Phi(^{71}{\rm Ga}) & = & (70.8 R({\rm pp}) + 3.1 R({\rm pep})
+ 35.8 R({^{7}\rm Be}) +\\ \nonumber
& & + 13.8 R(^{8}{\rm B}) + 7.9)\,{\rm SNU}
\end{eqnarray}
\end{enumerate}

In the above equations the quantities $R$ denote the
reaction rates (or low--energy astrophysical S--factors)
of eqs.~(1)--(3) normalized to the currently
accepted values given in (12). We want to emphasize that
eqs.~(1)--(6) can only give a semiquantitative understanding
of the dependence of the solar neutrino fluxes, because
the actual variations of the calculated neutrino fluxes are
determined by the coupled partial differential equations
of stellar evolution and the boundary conditions (48).
However, in many cases the formulae (1)--(6) can give a first
approximation of the dependence of the solar neutrino
fluxes on the astrophysical S--factors.

Using the new input rates discussed in the foregoing section
we can use the above formula to recalculate the neutrino
input fluxes from the individual neutrino fluxes as well
as the changes in the predicted rate for the neutrino
experiments. The pp neutrino flux is almost unchanged,
the $^{7}$Be and $^{8}$B neutrino fluxes increase by 1\%
and 11\%, respectively, when compared them to the
currently accepted values (12). The enhancement of the $^{8}$B neutrino flux
is due to the higher astrophysical S--factor
for the reaction $^{7}$Be(p,$\gamma$)$^{8}$B which is the
most uncertain nuclear reaction rate with respect
to the solar neutrino problem. With the new nuclear reaction
rates we get the following predictions for the neutrino fluxes
enforcing the solar neutrino problem to some extend:
\begin{enumerate}
\item $^{37}$Cl experiment: $\Phi(^{37}{\rm Cl}) = 8.7$\,SNU\\
(instead of 8.0\,SNU (12)).
\item Kamiokande experiment: $\frac{\Phi({\rm Kamiokande (Observation)})}
{\Phi({\rm Kamiokande (Prediction)})} = 0.42$\\
(instead of 0.47 (12)).
\item $^{71}$Ga experiments: $\Phi(^{71}{\rm Ga}) = 133.6$\,SNU\\
(instead of 131.5\,SNU (12)).
\end{enumerate}

We can also use the above formulae to use an astrophysical
S--factor for the reaction $^{7}$Be(p,$\gamma$)$^{8}$B determined
from the recent experiment at RIKEN measuring the
low--energy
$^{8}$B + $^{208}$Pb $\rightarrow$ p + $^{7}$Be + $^{208}$Pb
Coulomb dissociation cross section (48). Taking
the E1 and E2 contributions into account the astrophysical
S--factor for the reaction $^{7}$Be(p,$\gamma$)$^{8}$B is only
55\% (49) of the current accepted value (12). In this case we obtain
\begin{enumerate}
\item $^{37}$Cl experiment: $\Phi(^{37}{\rm Cl}) = 5.2$\,SNU\\
(instead of 8.0\,SNU (12)).
\item Kamiokande experiment: $\frac{\Phi({\rm Kamiokande (Observation)})}
{\Phi({\rm Kamiokande (Prediction)})} = 0.85$\\
(instead of 0.47 (12)).
\item $^{71}$Ga experiments: $\Phi(^{71}{\rm Ga}) = 123.9$\,SNU\\
(instead of 131.5\,SNU (12)).
\end{enumerate}

This brings the predicted neutrino flux of the high--energy
neutrinos within the experimental errors of the Kamiokande experiment
shown in Table 1.
However, the predicted
neutrino flux of the $^{37}$Cl experiment of 5.2\,SNU is
still outside the experimental errors of this experiment.
This result is also approximately obtained by a much more sophisticated
method using 1000 solar models with modified fluxes in a
Monte Carlo simulation (50). None of these considered models lie
within 3$\sigma$ measurement errors of the $^{37}$Cl--experiment.
This means that at the present time it is still not possible
to reconcile the $^{37}$Cl and Kamiokande
experiments within the framework of the SSM.

Electron screening plays an important role in low--energy
laboratory measurements
of nuclear reactions as well as in the solar plasma. However,
the physics behind the screening effects for these two cases
is quite different. In laboratory experiments the bound electrons are
inevitably present in the target and shield part of the
barrier of the bare nuclei, whereas in the plasma the
electrons occupy continuum states. At present there is a significant,
unexplained discrepancy between the shielding effect in
laboratory experiments and the theoretical predictions
for the screening energies (the effective energy increase
through screening) which are to low up to a factor of about 2
when compared to experiment (7), (51). However, the knowledge
of screening effects is absolutely necessary in order
to make reliable extrapolations of experimentally
determined astrophysical S--factors to the thermonuclear
energies of the sun. The above discussed discrepancy which
may also be due to some unknown nuclear effect raises
doubts about the correctness of these extrapolations
to thermonuclear energies.
\begin{center}
{\large {\bf SUMMARY}}
\end{center}
\vspace{4pt}
The solar neutrino problem has persisted pertinaciously solutions
within standard physics. In this respect the nuclear
reaction rates obtained from laboratory efforts by measuring
solar nuclear reactions in earth--bound accelerator
experiments or theoretical analyses are no exceptions.
The calculated reaction rates which are discussed in this work
even enforces the solar neutrino problem to some extend, because
they favour the higher astrophysical S--factor for
the reaction $^{7}$Be(p,$\gamma$)$^{8}$B. Even if
the reaction rate for this reaction is changed drastically
as measured by the recent experiment at RIKEN,
one can reproduce the Kamiokande experiment within
the experimental errors, but it seems not to be
possible to obtain consistency with the $^{37}$Cl experiment.
At the present time there exists stil a great uncertainty in
the extrapolation of the astrophysical S--factors to
thermonuclear energies due to the incorrect
treatment of screening effects or some other unknown nuclear
effect.
\vspace{6pt}
\begin{center}
{\large {\bf ACKNOWLEDGEMENTS}}
\end{center}
\vspace{4pt}

We want to thank the Fonds zur Fonds zur Fšrderung der
wissenschaftlichen Forschung in \"Osterreich (FWF),
Hochschul\-ju\-bi\-l\"a\-ums\-stif\-tung
and the Deutsche Forschungsgemeinschaft (DFG) for their support. We are also
indebted
to our coworkers H.~Abele, K.~Gr\"un, H.~Krauss and P.~Mohr
for their assistance.
\vspace{6pt}
\begin{center}
{\large {\bf REFERENCES}}
\end{center}
\vspace{4pt}

{\small
\begin{enumerate}
\parindent0em
\item Wildermuth, K., and Tang, Y.~C., {\it A Unified Theory of the
Nucleus, Clustering Phenomena in Nuclei,
Vol.~3\/}, Braunschweig: Vieweg \& Sohn, 1977.
\item Baye, D., and Descouvemont, P., in {\it Proc.~5th Int.~Conf.~on
Clustering Aspects in Nuclear and Subnuclear Systems, Kyoto, Japan 1988\/},
{\it J.~Phys.~Soc.~Jpn.\/} {\bf 58}, Suppl.~103 (1989).
\item Langanke, K., in {\it Nuclear Astrophysics\/},
ed.~by J.W.~Negele, E.~Vogt, New York: Plenum Press, 1986, p.~223.
\item Langanke, K., and Friedrich, H., in {\it Advances in Nuclear Physics,
Vol.~17\/}, ed.~by M.~Lozano, M.~I.~Gallardo, J.~M.~Arias, Berlin:
Springer--Verlag,
1988, p.~241.
\item Oberhummer, H., and Staudt G., in {\it Nuclei in the Cosmos\/},
ed.~by H.~Oberhummer, Berlin: Springer--Verlag, 1991, p.~28.
\item Langanke, K., in {\it Nuclei in the Cosmos\/},
ed.~by H.~Oberhummer, Berlin: Springer--Verlag, 1991, p.~61.
\item Langanke, K., and Barnes, C.~A., in
{\it VI Swieca Summer School, Sao Paulo Univ., Brazil\/},
ed.~by D.~Menzenes, F.~Navarra, Singapore: World Scientific, 1993, in press.
\item Kobos, A.~M., Brown, B.~A., Lindsay, R., and Satchler, G.~R.,
{\it Nucl.~Phys.~A\/} {\bf 425}, 205 (1984).
\item Abele, H., and Staudt, G., {\it Phys.~Rev.~C\/} {\bf 47}, 742 (1993).
\item Oberhummer, H., {\it Kerne und Sterne\/}, Leipzig: Johann
Ambrosius Barth, 1993.
\item Bahcall, J.~N., {\it Neutrino Astrophysics\/}, Cambridge:
Cambridge Univ.~Press, 1989.
\item Bahcall, J.~N., and Pinsonneault, M.~H.,
{\it Rev.~Mod.~Phys.\/} {\bf 64}, 885 (1992).
\item Wang, N.~M., Novatskii, V.~M., Osetinskii, G.~M., Chien, N.~K.,
and Chepurchenko, I.~A.,
{\it Sov.~J.~Nucl.~Phys.\/} {\bf 3}, 777 (1966).
\item Bacher A.~D., and Tombrello T.~A., quoted by T.~A.~Tombrello,
in {\it  Nuclear Research with low--energy accelerators\/}, ed.~by
J.~B. Marion and D.~M.~van Patter, New York: Academic Press, 1967, p.~195.
\item Dwarakanath, M.~R., and Winkler, H.,
{\it Phys.~Rev.~C\/} {\bf 4}, 1532 (1971).
\item Dwarakanath, M.~R.,
{\it Phys.~Rev.~C\/} {\bf 9}, 805 (1974).
\item Krauss, A., Becker, H.~W., Trautvetter, H.~P., and Rolfs, C.,
{\it Nucl.~Phys.~A\/} {\bf 467}, 273 (1987).
\item Mc Donald, A.~B., Alexander, T.~K., Beene, J.~E., and Mak, H.,~B.,
{\it Nucl. Phys.~A\/} {\bf 288}, 529 (1977).
\item Fetisov, V.~N., and Kopysov, Y.~S.,
{\it Phys.~Lett.~B\/} {\bf 40}, 602 (1972);
{\it Nucl. Phys.~A\/} {\bf 239}, 511 (1975).
\item Fowler, W.~A.,
{\it Nature\/} {\bf 238}, 24 (1972).
\item Rolfs, C., Rodney, W.~S., and Trautvetter, H.~P.,
{\it Rep.~Prog.~Phys.\/} {\bf 50}, 233 (1987).
\item Typel, S., Bl\"uge, G., Langanke K., and Fowler, W.~A.,
{\it Z.~Phys.~A\/} {\bf 339}, 249 (1991).
\item Krauss, H., GrŸn, K., Herndl, H., Rauscher, T., Oberhummer, H.,
Abele, H., Mohr, P., Zwiebel, R., Staudt, G., Denker, A., Wolf, G.,
and Hammer H., in {\it Proceedings
of the 2nd International Symposium on Nuclear Astrophysics,
Nuclei in the Cosmos, Karlsruhe\/}, ed.~by F.~KŠppeler und K.~Wisshak,
Bristol: IOP Publishing, 1993, p.~393.
\item Parker, P.~D., and Kavanagh, R.~W.,
{\it Phys.~Rev.~C\/} {\bf 131}, 2578 (1974).
\item Nagatani, K.,~Dwarakanath, M.~R., and Ashery~D.,
{\it Nucl.~Phys.~A\/} {\bf 128}, 325 (1969).
\item Kr\"ahwinkel, H., Becker, H.~W., Buchmann, L., G\"orres, J.,
Kettner, K.~U., Kieser, W.~E., Santo, R., Schmalbrock, P.,
Trautvetter, H.~P., Vlieks, A., Rolfs, C., Hammer, J.~W.,
Azuma, R.~E., and Rodney, W.~S.,
{\it Z.~Phys.~A\/} {\bf 304}, 307 (1982).
\item  Osborne, L., Barnes, C.~A., Kavanagh, R.~W., Kremer, R.~M., Mathews
G.~J.,
Zyskind, J.~L., Parker, P.~D., and Howard, A.~J.,
{\it Nucl.~Phys.~A\/} {\bf 419}, 115 (1984).
\item Alexander, T.~K., Ball G.~C., Lennard, W.~N., Geissel, H., and
Malc, H.-B.,
{\it Nucl.~Phys.~A\/} {\bf 427}, 526 (1984).
\item Hilgemeier, M., Becker, H.~W., Rolfs, C., Trautvetter, H.~P.,
and Hammer, J.~W.,
{\it Z.~Phys.~A\/} {\bf 329}, 243 (1988).
\item Robertson, H.~G.~H., Dyer, P., Bowles, T.~J., Brown, R.~E.,
Jarmie, N., Maggiore, C.~J., and Austin, S.~M.,
{\it Phys.~Rev.~C\/} {\bf 27}, 11 (1983).
\item Volk, H., Kr\"ahwinkel H., Santo R., and Wallek L.,
{\it Z.~Phys.~A\/} {\bf 310}, 91 (1983).
\item Tombrello, T.~A., and Parker, P.~D.,
{\it Phys.~Rev.~C\/} {\bf 131}, 2582 (1963).
\item Kim, B.~T., Izumoto, T., and Nagataki, K.,
{\it Phys.~Rev.~C\/} {\bf 23}, 33 (1981).
\item Buck, B., Baldock, R.~A., and Rubio, J.~A.,
{\it J.~Phys.~G\/} {\bf 11}, L11 (1985).
\item Waliser, H., Kanada, H., and Tang, Y.~C.,
{\it Nucl.~Phys.~A\/} {\bf 419}, 133 (1984).
\item Kajino, T., and Arima, A.,
{\it Phys.~Rev.~Lett.\/} {\bf 52}, 739 (1986).
\item Mertelmeier, T., and Hofmann, H.~M.,
{\it Nucl.~Phys.~A\/} {\bf 459}, 387 (1986).
\item Langanke, K.,
{\it Nucl.~Phys.~A\/} {\bf 457}, 351 (1986).
\item Kajino, T., {\it Nucl.~Phys.~A\/} {\bf 460}, 559 (1986).
\item Liu, Q.~K.~K., Kanada, H., and Tang, Y.~F.,
{\it Phys.~Rev.~C\/} {\bf 33}, 1561 (1986).
\item Mohr, P., Abele, H., Zwiebel, R., Staudt, G., Krauss, H., Oberhummer, H.,
Denker, A., Hammer, J.~W., and Wolf, G.,
{\it Phys.~Rev.~C\/} {\bf 48}, 1420 (1993).
\item Kavanagh, R.~W.,
{\it Nucl.~Phys.\/} {\bf 15}, 411 (1986);
{}~Kavanagh, R.~W., Tombrello, T.~A., Mosher, J.~M., and Goosman, D.~R.,
{\it Bull.~Am.~Phys.~Soc.\/} {\bf 14}, 1209 (1969).
\item Filipone, B.~W., Elwyn, A.~J., Davids, C.~N., and Koethe, D.~D.,
{\it Phys.~Rev. Lett.\/} {\bf 50}, 412 (1983); {\it Phys.~Rev.~C\/} {\bf 28},
2222 (1983).
\item Descouvement, P., and Baye, D.,
{\it Nucl.~Phys.~A\/} {\bf 459}, 387 (1986).
\item Johnson, C.~W., Kolbe, E., Koonin, S.~E., and Langanke, K.,
{\it Ap.~J.\/} {\bf 392}, 320 (1992).
\item Descouvement, P., and Baye, D.,
{\it Nucl.~Phys.~A\/}, in press.
\item Krauss, H., Gr\"un, K., Rauscher, T., and Oberhummer, H.,
{\it Ann.~Phys.\/} {\bf 2}, 256 (1993).
\item Motobayashi, T., et al., RIKEN preprint Rikkyo RUP 94--2, 1994,
{\it Phys.~Rev. Lett.\/}, submitted.
\item Langanke, K., Shoppa, T.~D., Caltech preprint MAP--168, 1994.
\item Bahcall, J.~N., and Bethe, H.~A.,
{\it Phys.~Rev.~D\/} {\bf 47}, 1298 (1993).
\item Leclerq--Willain, C., Demeur, M., Azrak, Z., Wauters, L.,
Universite Libre de Bruxelles, Physique 229 preprint, PNT/13/93, 1993.
\end{enumerate}}
\end{document}